\documentstyle[12pt]{article}
\begin{document}
\begin{center}
{\Large \bf A NOTE ON THE $\gamma \gamma\rightarrow \pi^0\pi^0$ REACTION IN
 THE $1/N$
 EXPANSION OF  $\chi PT$}
\vskip 1.0cm
{\large \ Antonio DOBADO \footnote{E-mail: dobado@cernvm.cern.ch} }   \\
 Departamento de F\'{\i}sica Te\'orica  \\
 Universidad Complutense de Madrid\\
 28040 Madrid, Spain \\
\vskip 0.5cm
and
\vskip 0.5cm
{\large \ John MORALES \footnote{ On leave of absence from Centro Internacional
de F\'{\i}sica, Colombia} }   \\
 Departamento de F\'{\i}sica Te\'orica  \\
 Universidad Aut\'onoma de Madrid\\
 28049 Madrid, Spain \\
\vskip 1.0cm
{\bf \large September 1995}
\vskip 1.0cm
\begin{abstract}
In this work we present the results of a complete calculation of the  $\gamma
 \gamma\rightarrow
\pi^0\pi^0$ amplitude to  leading order in the large $N$ approximation ($N$
being the number
of Goldstone bosons) up to order $m^2_{\pi}/F^2$. The amplitude turns to be
proportional to
that of
$\pi^+\pi^-\rightarrow \pi^0\pi^0$. In spite of the fact that this
factorization property cannot
hold in general (as it  was recently pointed out by Morgan and Penington), it
appears here since
in the large $N$ limit only the $I=J=0$ channel contributes to the $\gamma
\gamma\rightarrow
\pi^0\pi^0$ reaction. Moreover it seems to be a reasonable approximation in
 this case since it
is possible to reproduce, as a prediction, the experimental data starting from
a one-parameter fit of the $\pi\pi$ scattering data.

\end{abstract}
\end{center}
\vskip 1.0cm

\newpage

\section{Introduction}
\hspace*{12pt}
In the last years a great deal of work has been devoted to the $\gamma
\gamma\rightarrow \pi^0\pi^0$
reaction. The reasons for that are the experimental data available from the
the Crystal Ball
Collaboration [1] and that this process, being a pure finite one-loop effect,
 is a very good test
 for
the modern Chiral Perturbation Theory ($\chi PT$) [2] which successfully
describes many
low-energy hadronic reactions. However,
 for $\gamma \gamma\rightarrow \pi^0\pi^0$, the one-loop
prediction [3] disagrees with the  experimental data even near threshold.
 This is due to the  pion rescattering which is very strong in the $I=J=0$
channel.
Nevertheless it has been shown
that the use of dispersion relation techniques [4] can provide a good
description of this
 reaction
 and reconcile data with $\chi PT$ [5]. More recently it has been shown
that a two loop-computation produces a cross-section that agrees rather well
with the available data and with dispersion theoretical calculations even
substantially above threshold [6].

Another completely different approach that has been proposed for this reaction
 by
Im [7] is the so called large
 $N$ expansion ($N$ being the number of Nambu-Goldstone bosons). This
 method was first used
for elastic pion scattering in [8] in the context of the Linear Sigma Model
(LSM)
 and in [9] in
the Non Linear Sigma Model (NLSM). Pion mass effects have also been
 included in [10] for the LSM and in [11] for the NLSM.
In this last work it was shown how it is possible to fit the $J=0$ pion
scattering
phase-shifts up to very high energies.

Nevertheless  Im's work has been strongly critizised by Pennington and Morgan
 in [12] because some of his assumptions are only true for point interactions
which
 is not the case of pions. Since  Im's computation is not in fact a
 complete large $N$ computation, these assumptions cannot be supported and
 thus we consider  Pennington and Morgan criticisms appropriate. In this
 note we clarify this issue by making a complete large $N$ computation of the
 $\gamma \gamma\rightarrow \pi^0\pi^0$ reaction. Our main conclusion is that
at the end the original Im's result is correct at leading
order of the large $N$ expansion and to first order in
 $m_{\pi}^2/F^2$ so both Pennington and Morgan on one side as Im
 on the other,  are simultaneously right. In addition we will show how the
large $N$
 approximation provides a very good description of the Crystal Ball
 experimental data.

\section{The Chiral Lagrangian in Standard  $S^3$ Coordinates}
\hspace*{12pt}

As it is well known, the low-energy interactions of pions and photons can
 be described by a $U(1)_{em}$ gauged NLSM in the context of
modern $\chi PT$. This NLSM is based in the coset space
$SU(2)_L \times SU(2)_R/SU(2)_{L+R}=O(4)/O(3)= S^3 = SU(2)$. In order to define
 the large $N$ limit we will extend the coset space to
  $O(N+1)/O(N)= S^N$ and we will define the corresponding lagrangian as
follows. First we
start from the LSM
\begin{eqnarray}
{\cal L} = \frac{1}{2} {\partial}_{\mu}{\phi}^{T} {\partial}^{\mu}{\phi}
 + \sqrt{N}Fm^{2} \sigma
\end{eqnarray}
where ${\phi}^{T}=({\pi}_{1},{\pi}_{1}^{'},{\pi}_{2},{\pi}_{2}^{'},..,
{\pi}_{k}
,{\pi}_{k}^{'},{\pi}_{N},{\sigma})$
which is $O(N+1)$ invariant if it was not for the last piece that explicitly
breaks
 this symmetry
to $O(N)$ (with  $N=2k+1$). This contribution will provide the mass term for
the pions and other
 interactions
 forced by the chiral
 symmetry. With ${\pi}_{0}={\pi}_{N}$
the above lagrangian can be written as:
\begin{eqnarray}
{\cal L} = \frac{1}{2} {\partial}_{\mu}{\pi}_{a} {\partial}^{\mu}{\pi}^{a}
 + \frac{1}{2} {\partial}_{\mu}{\pi}_{a}^{'} {\partial}^{\mu}{\pi}^{'a} +
\frac{1}{2} {\partial}_{\mu}{\pi}^{0} {\partial}^{\mu}{\pi}^{0} + \frac{1}{2}
 {\partial}_{\mu}{\sigma} {\partial}^{\mu}{\sigma} + \sqrt{N}Fm^{2} {\sigma}
\end{eqnarray}
with $a=1,2,..,k $, or, in terms of the complex charged fields
${\varphi}^{a}=({\pi}^{a}+i{\pi}^{'a})/\sqrt{2}$, as
\begin{eqnarray}
{\cal L} = {\partial}_{\mu}{\varphi}_{a}^{*} {\partial}^{\mu}{\varphi}^{a} +
\frac{1}{2} {\partial}_{\mu}{\pi}^{0} {\partial}^{\mu}{\pi}^{0} + \frac{1}{2}
{\partial}_{\mu}{\sigma} {\partial}^{\mu}{\sigma} + \sqrt{N}Fm^{2} \sigma
\end{eqnarray}
The electromagnetic interactions are included by introducing the corresponding
covariant derivative
\begin{eqnarray}
{\partial}_{\mu}{\varphi}^{a} \rightarrow D_{\mu}{\varphi}^{a} \equiv
 {\partial}_{\mu}{\varphi}^{a} + ieA_{\mu}{\varphi}^{a}
\end{eqnarray}
and adding the photon kinetic term
\begin{eqnarray}
{\cal L}_{g} & = & {\partial}_{\mu}{\varphi}_{a}^{*} {\partial}^{\mu}
{\varphi}^{a} + \frac{1}{2}{\partial}_{\mu}{\pi}^{0} {\partial}^{\mu}{\pi}^{0}
+
 \frac{1}{2} {\partial}_{\mu}{\sigma} {\partial}^{\mu}{\sigma} +
\sqrt{N}Fm^{2} {\sigma}
\nonumber \\
& + & e^{2}A^{2}{\varphi}_{a}^{*}{\varphi}^{a} + ieA_{\mu} {\partial}
_{\mu}{\varphi}_{a}^{*}{\varphi}^{a} + h.c. - \frac{1}{4}F_{{\mu}{\nu}}
F^{{\mu}{\nu}}
\end{eqnarray}
where $F_{{\mu}{\nu}}={\partial}_{\mu}A_{\nu}-{\partial}
_{\mu}A_{\nu}$. Now we obtain the NLSM
describing the low-energy pion dynamics by setting ${\phi}^{T}{\phi}=
NF^{2}=f_{\pi}^2$ to force the
 system to live in the $ S^{N}$ sphere i.e.
\begin{eqnarray}
{\pi}^{2}+{\sigma}^{2}=NF^{2}
\end{eqnarray}
where ${\pi}^{2}=\sum_{a=1}^k({{\pi}^{a}}^{2}+{{\pi'}^{a}}^{2})$. Thus we have
${\sigma}=(NF^{2}-{\pi}^{2})^{1/2}$

and the lagrangian of the $U(1)_{em}$ gauged NLSM becomes
\begin{eqnarray}
{\cal L}_{g} & = & {\partial}_{\mu}{\varphi}^{+} {\partial}^{\mu}{\varphi}
 + \frac{1}{2} {\partial}_{\mu}{\pi}^{0} {\partial}^{\mu}{\pi}^{0} +
\frac{1}{2}
\frac{\left({\partial}_{\mu}{\varphi}^{+}{\varphi} + {\varphi}^{+}
 {\partial}^{\mu}{\varphi} + {\pi}^{0} {\partial}_{\mu}{\pi}^{0}\right)^{2}}
{NF^{2}-2{\varphi}^{+}{\varphi}}
\nonumber \\
& + & NF^{2}m^{2} \sqrt{1-\frac{2{\varphi}^{+}{\varphi} + {\pi}_{0}^{2}}
{NF^{2}}}
+ e^{2}A^{2}{\varphi}^{+}{\varphi} + ieA_{\mu} {\partial}^{\mu}{\varphi}^{+}
{\varphi} + h.c.
\nonumber \\
& - & \frac{1}{4}F_{{\mu}{\nu}}F^{{\mu}{\nu}}
\end{eqnarray}
with ${\varphi}^{T}=({\varphi}_{1},...,{\varphi}_{k})$.

This way to write
the lagrangian describing the low-energy pion dynamics corresponds to a
particular coordinate choice on the coset space $S^N$ but in principle any
other
 parametrization connected with this one through an analytical change of
 coordinates on the coset manifold
will yield  the same $S$ matrix elements
 (but not necessarily  the same  Green functions) [13]. In
 particular, for the
 $N=3$ case
a very popular coordinate choice consists in defining the following $SU(2)=
S^3$
 matrix:
\begin{eqnarray}
{\Sigma} = e^{\frac{i{\tau}^{a}{\omega}^{a}(x)}{f_{\pi}}}
\end{eqnarray}
The change of coordinates ${\phi}^{i}
={\phi}^{i}({\omega})$, with  $i=1,2,3$ relating both parametrizations of the
 coset space is given by
$f_{\pi}{\Sigma}={\phi}^{4}+i{\phi}^{i}
{\tau}^{i}$ with ${\phi}^{4}=\sqrt{f_{\pi}^{2}-{\phi}^{2}}$ where ${\phi}^{2}
=\sum_{i=1}^{3}{{\phi}^{i}}^2$ with ${\phi}^{T}=
[{\phi}_{1},{\phi}_{2},{\phi}_{3},{\phi}_{4}]=
[{\pi}_{1},{\pi}_{1}',{\pi}_{0},\sigma]$. The above lagrangian written in
terms of these new
 coordinates reads

\begin{eqnarray}
{\cal L} = \frac{f_{\pi}^{2}}{4}Tr D_{\mu}{\Sigma}\left(D^{\mu}{\Sigma}\right)^
{+} + \frac{f_{\pi}^{2}}{4}Tr m^{2}\left({\Sigma} + {\Sigma}^{+}\right)
 - \frac{1}{4}F_{{\mu}{\nu}}F^{{\mu}{\nu}}
\end{eqnarray}
where
\begin{eqnarray}
D_{\mu}{\Sigma} & = & {\partial}_{\mu}{\Sigma} + ie \left[Q,{\Sigma}
 \right]A_{\mu}
\nonumber \\
Q & = & diag \left(2/3,-1/3 \right)
\end{eqnarray}
As it was commented above, both lagrangians yield the same $S$ matrix elements
and the same observables like phase-shifts, cross sections, etc. However, the
Green
functions could be different. Indeed the computations can be much easier using
one set of
coordinates than another since the Feynman rules and the diagrams contributing
to some
given process will be different in general.  In particular, the computation of
the one-loop
$\gamma \gamma\rightarrow \pi^0\pi^0$ amplitude is simpler using the $\phi$
 standard $S^3$
coordinates
than the $\Sigma$ (chiral) ones, which were used in [3]
\section{The one-loop scattering amplitude}
\hspace*{12pt}
The ${\gamma}{\gamma} \rightarrow {\pi}{\pi}$ transition amplitude is defined
 as
\begin{eqnarray}
{\cal A} \left({\gamma}{\gamma} \rightarrow {\pi}^{0}{\pi}^{0},{\pi}^{+}
{\pi}^{-} \right) = {\epsilon}_{1\mu} \left(k_{1} \right){\epsilon}_{2\nu}
 \left(k_{2} \right){\cal M}_{{\pi}^{0}{\pi}^{0},{\pi}^{+}{\pi}^{-}}^{{\mu}
{\nu}}
\end{eqnarray}
where ${\epsilon}_{1\mu},{\epsilon}_{2\nu}$ and $k_{1},k_{2}$ are the
 polarization vectors
and the momenta of the two photons.

The two one-loop diagrams for the process ${\gamma}{\gamma} \rightarrow
 {\pi}^{0}{\pi}^{0}$
using standard $S^3$ coordinates are shown in Fig.1.a. Note that when this
 computation is
 done with chiral coordinates there
are two more Feynman diagrams [3]. In our case, using dimensional
regularization,
 the diagrams are
 given by:

\begin{eqnarray}
{\cal N}^{a}_{{\mu}{\nu}} & = & \frac{2e^{2}}{ \left(NF^{2} \right)^{2}}
 \left(s - m_{\pi}^{2} \right) \int d\tilde{\ell} \frac{g_{{\mu}{\nu}}}
{ \left[ \left({\ell}+k_{1} \right)^{2}-m_{\pi}^{2} \right] \left[ \left({\ell}
-k_{2}\right)^{2}-m_{\pi}^{2} \right]}
\nonumber \\
{\cal N}^{b}_{{\mu}{\nu}} & = & \frac{-e^{2}}{ \left(NF^{2} \right)^{2}}
\left(s - m_{\pi}^{2} \right) \int d\tilde{\ell} \frac{ \left(2{\ell}+k_{1}
 \right)_{\mu} \left(2{\ell}-k_{2} \right)_{\nu}}{ \left({\ell}^{2}-m_{\pi}^{2}
 \right) \left[ \left({\ell}+k_{1} \right)^{2}-m_{\pi}^{2} \right]
 \left[ \left({\ell}-k_{2} \right)^{2}-m_{\pi}^{2} \right]}
\nonumber \\
& + & \left[\left(k_{1},{\mu} \right) \longleftrightarrow \left(k_{2},{\nu}
\right)
\right]
\end{eqnarray}
with
\begin{eqnarray}
\int d\tilde{\ell} \equiv {\mu}^{\epsilon}
\int \frac{d^{4-\epsilon}{\ell}}{\left(2{\pi}\right)^{4-\epsilon}}
\end{eqnarray}

It is not difficult to see that the above  ${\cal N}^{a}_{{\mu}{\nu}}$
 and ${\cal N}^{b}_{{\mu}{\nu}}$ correspond respectively to the terms
${\cal M}^{a}_{{\mu}
{\nu}}+{\cal M}^{c}_{{\mu}{\nu}}$ and ${\cal M}^{b}_{{\mu}{\nu}}+{\cal M}^{d}
_{{\mu}{\nu}}$ of the second reference in [3].
The complete matrix element is given by

\begin{eqnarray}
{\cal M}_{{\mu}{\nu}} = \frac{2e^{2}}{ \left(NF^{2} \right)^{2}}
 \left(s - m_{\pi}^{2} \right) \int d\tilde{\ell} \frac{g_{{\mu}{\nu}}
 \left(l^{2}-m_{\pi}^{2}\right)- \left(2{\ell}+k_{1} \right)_{\mu}
 \left(2{\ell}-k_{2} \right)_{\nu}}{ \left({\ell}^{2}-m_{\pi}^{2} \right)
 \left[ \left({\ell}+k_{1} \right)^{2}-m_{\pi}^{2} \right]
\left[ \left({\ell}-k_{2} \right)^{2}-m_{\pi}^{2} \right]}
\end{eqnarray}
which is proportional to the $\chi PT$ lowest order $\pi^+\pi^-\rightarrow
\pi^0\pi^0$ amplitude
given by $(s - m_{\pi}^{2})/f_{\pi}^2$.

As it is well known, the integral above is finite and therefore no
renormalization is needed.
This is because there is no four derivative term in the gauged NLSM
contributing to
$\gamma \gamma\rightarrow \pi^0\pi^0$. Therefore there is
no any possible counterterm to absorb the divergences appearing in the
one-loop computation and
thus the consistency of $\chi PT$ renders this one-loop amplitude completely
finite. The
result is given by
\begin{eqnarray}
{\cal M}_{{\mu}{\nu}} & = & \frac{2e^{2}}{ \left(NF^{2} \right)^{2}}
 \left(s - m_{\pi}^{2} \right) \left( \frac{-i}{16{\pi}^{2}} \right)
 \left(g_{{\mu}{\nu}}- \frac{k_{2{\mu}}k_{1{\nu}}}{k_{1}k_{2}} \right) \left
[1+ \frac{m_{\pi}^{2}}{s} \left\{{\ln}Q_{\pi}-i{\pi} \right\}^{2} \right]
\nonumber \\
Q_{\pi} & = & \frac{1+ \sqrt{1- \frac{4m_{\pi}^{2}}{s}}}
{1- \sqrt{1- \frac{4m_{\pi}^{2}}{s}}}
\end{eqnarray}
{}From this amplitude one can calculate the total cross-section which is found
to be

\begin{eqnarray}
{\sigma} \left({\gamma}{\gamma} \rightarrow {\pi}^{0}{\pi}^{0} \right)
 = \frac{{\alpha}^{2}}{8{\pi}^{2}} \sqrt{1- \frac{4m_{\pi}^{2}}{s}}
 \left[1+ \frac{m_{\pi}^{2}}{s}f \left(s \right) \right]{\sigma}_0
 \left({\pi}^{+}{\pi}^{-} \rightarrow {\pi}^{0}{\pi}^{0} \right)
\end{eqnarray}

where

\begin{eqnarray}
{\sigma}_0 \left({\pi}^{+}{\pi}^{-} \rightarrow {\pi}^{0}{\pi}^{0} \right) =
\frac{1}{32{\pi} \left(NF^{2} \right)^{2}} \frac{ \left(s-m_{\pi}^{2}
 \right)^{2}}{s}
\end{eqnarray}
which is the lowest order ${\pi}^{+}{\pi}^{-} \rightarrow {\pi}^{0}{\pi}^{0} $
scattering
 cross-section
and

\begin{eqnarray}
f \left(s \right) = 2 \left[{\ln}^{2} \left( \frac{1+
 \sqrt{1- \frac{4m_{\pi}^{2}}{s}}}{1- \sqrt{1- \frac{4m_{\pi}^{2}}{s}}} \right)
-{\pi}^{2} \right]+ \frac{m_{\pi}^{2}}{s} \left[{\ln}^{2}
 \left( \frac{1+ \sqrt{1- \frac{4m_{\pi}^{2}}{s}}}{1- \sqrt{1- \frac{4m_{\pi}
^{2}}{s}}} \right)+{\pi}^{2} \right]^{2}
\end{eqnarray}

Thus,
from the above equations one can realize that, as it was remarked in the
second reference in [3],
the lowest order
 ${\pi}^{+}{\pi}^{-} \rightarrow {\pi}^{0}{\pi}^{0} $
amplitude and cross-section factorize the  one-loop
${\gamma}{\gamma} \rightarrow {\pi}^{0}{\pi}^{0}$ ones.

{}From the diagrams in Fig.1. it is also possible to compute
the reaction ${\gamma}{\gamma}\rightarrow{\pi}^{+}{\pi}^{-}$. However, in this
case
we have  a tree level contribution (Fig.1.b) and, in addition, we have to
consider
at the one-loop level (Fig.1.c) the  contribution coming from the pion
wave-function
renormalization (Fig.1.d).
In the reference frame where the photon polarization vectors ${\epsilon}_{1},
{\epsilon}_{2}$ satisfy
\begin{eqnarray}
k_{1} \cdot {\epsilon}_{1} = k_{1} \cdot {\epsilon}_{2} = k_{2} \cdot
 {\epsilon}_{1} = k_{2} \cdot {\epsilon}_{2} = 0
\end{eqnarray}
the amplitude at  tree level is given by
\begin{eqnarray}
{\cal A}_{ \left({\gamma}{\gamma} \rightarrow {\pi}^{+}{\pi}^{-}
 \right)}^{tree}=2ie^{2} \left\{{\epsilon}_{1} \cdot {\epsilon}_{2}
- \frac{p^{+} \cdot {\epsilon}_{1} p^{-} \cdot {\epsilon}_{2}}{k_{1}
 \cdot p^{+}}- \frac{p^{-} \cdot {\epsilon}_{1} p^{+} \cdot {\epsilon}_{2}}
{k_{2} \cdot p^{+}} \right\}
\end{eqnarray}
where $k_{1},k_{2}$ are the photon momenta and $p^{+},p^{-}$
are the  ${\pi}^{+},{\pi}^{-}$ outgoing momenta. Now including the one-loop
corrections.
\begin{eqnarray}
{\cal A}_{ \left({\gamma}{\gamma} \rightarrow {\pi}^{+}{\pi}^{-} \right)}
^{one-loop} & = & \frac{-ie^{2}}{16{\pi}^{2}NF^{2}}
\left\{{\epsilon}_{1} \cdot {\epsilon}_{2}
\left(s+m_{\pi}^{2}{\ln}^{2}Q_{\pi}
\right) \right.
\nonumber \\
 & + & \left. 2A_{0}
\left(m_{\pi}^{2}
\right)
\left[{\epsilon}_{1} \cdot {\epsilon}_{2}
- \frac{p^{+} \cdot {\epsilon}_{1} p^{-} \cdot {\epsilon}_{2}}
{k_{1} \cdot p^{+}}
- \frac{p^{-} \cdot {\epsilon}_{1} p^{+} \cdot {\epsilon}_{2}}
{k_{1} \cdot p^{-}}
\right]
\right\}
\end{eqnarray}
with
\begin{eqnarray}
A_{0} \left(m_{\pi}^{2} \right) & = & m_{\pi}^{2} \left({\bigtriangleup}
 - {\ln}m_{\pi}^{2} \right)
\nonumber \\
{\bigtriangleup} & = &\frac{2}{\epsilon}+\log{4\pi}-\gamma+1+{\ln}{\mu}^{2}
\end{eqnarray}
so that the full amplitude is
\begin{eqnarray}
{\cal A} \left({\gamma}{\gamma} \rightarrow {\pi}^{+}{\pi}^{-} \right)=
 \left[{\cal A}_{ \left({\gamma}{\gamma} \rightarrow {\pi}^{+}{\pi}^{-}
\right)}^{tree}+{\cal A}_{ \left({\gamma}{\gamma} \rightarrow
{\pi}^{+}{\pi}^{-}
 \right)}^{one-loop} \right]{\cal Z}_{\pi}
\end{eqnarray}
where the wave function renormalization constant is given by
\begin{equation}
{\cal Z}_{\pi}  =  1 + \frac{1}{16{\pi}^{2}NF^2}
A_{0} \left(m_{\pi}^{2} \right)
\end{equation}
so that we can write
\begin{eqnarray}
{\cal A} \left({\gamma}{\gamma} \rightarrow {\pi}^{+}{\pi}^{-} \right) =
2ie^{2}
 \left\{a{\epsilon}_{1} \cdot {\epsilon}_{2}- \frac{p^{+} \cdot {\epsilon}_{1}
 p^{-} \cdot {\epsilon}_{2}}{k_{1} \cdot p^{+}}- \frac{p^{-} \cdot
{\epsilon}_{1} p^{+} \cdot {\epsilon}_{2}}{k_{2} \cdot p^{+}}\right\}
\end{eqnarray}
where $a$ is given by
\begin{eqnarray}
a(s)=1- \frac{1}{16{\pi}^{2}NF^{2}} \left( \frac{s}{2} + \frac{m_{\pi}^{2}}{2}
 {\ln}^{2}Q_{\pi} \right)
\end{eqnarray}

\section{The large $N$ limit approximation}

In order to compute  the ${\gamma}{\gamma} \rightarrow {\pi}^{0}{\pi}^{0}$
 cross-section
at leading order in the $1/N$ expansion we go back to the lagrangian
 in eq.7. The pion
scattering amplitudes to order $1/N$ and the lowest order in
 $m^2_{\pi}/f^2_{\pi}$ were computed in [11]. The contributing
diagrams in this approximation are shown in Fig.2.a where the pion
 vertex also includes
the infinite set of counter terms needed to renormalize the bare result.
The renormalized amplitude can be written as

\begin{eqnarray}
T_{abcd}=
\delta_{ab}\delta_{cd}A\left(s\right)+
\delta_{ac}\delta_{bd}A\left(t\right)+
\delta_{ad}\delta_{bc}A\left(u\right)
\end{eqnarray}
where
\begin{eqnarray}
A\left(s\right)=
\frac{1}{NF^2}
\frac{G^{R}(s;\mu)}
{1-
\frac{sG^{R}(s;\mu)}{2\left(4\pi\right)^2F^2}
T(s;\mu)}
\left\{s-
\frac{m^2_{\pi}G^{R}(s;\mu)}
{1-
\frac{sG^{R}(s;\mu)}{2\left(4\pi\right)^2F^2}
T(s;\mu)}
\left[J^{R}(s;\mu)
-\frac{s}{\left(4\pi\right)^2F^2}\log{\frac{m_{\pi}^2}{\mu^2}}
\right]
\right\}
\end{eqnarray}
with
\begin{eqnarray}
T\left(s;\mu\right)\equiv
\sqrt{1-\frac{4m_{\pi}^2}{s}}
\log{
\frac{\sqrt{1-\frac{4m_{\pi}^2}{s}}-1}{\sqrt{1-\frac{4m_{\pi}^2}{s}}+1}
}-
\log{
\frac{m^2_{\pi}}{\mu^2}}
\end{eqnarray}
The $G^R(s;\mu)$ and   $J^R(s;\mu)$ are the generating
 functions of the counterterm couplings and they have a well defined dependence
on the $\mu$ scale, in order to ensure the $\mu$ independence of the whole
 amplitude. The isospin channels for the
model here considered  are  defined as [14]:
\begin{eqnarray}
T_{0} & = & NA\left(s\right)+A\left(t\right)+A\left(u\right)
\nonumber\\
T_{1} & = & A\left(t\right)-A\left(u\right)
\nonumber\\
T_{2} & = & A\left(t\right)+A\left(u\right)
\end{eqnarray}
and the partial waves are given by:
\begin{eqnarray}
a_{IJ}(s)=
\frac{1}{64\pi}
\int_{-1}^{1} d\left(\cos{\theta}\right)
T_{I}\left(s,\cos{\theta}\right)
P_{J}\left(\cos{\theta}\right)
\end{eqnarray}
As it is well known, the requirement of unitarity strongly constraints the
 possible behaviour of these partial waves. In particular, they should have a
cut  along the positive real axis from the threshold to infinity. The physical
amplitudes are the values just on the cut, and, in addition, the
condition of elastic unitarity:
\begin{equation}
Im  a_{IJ}  =
\sigma |a_{IJ}|^2
\end{equation}
(where $
\sigma  =  \sqrt{1-4m^2_{\pi}/s}$) has to be satisfied in the physical
region $s= E^2+i\epsilon$ and
$E^2  >  4m^2_{\pi}$, where $E$ is the total center of mass energy. This
equation is
exact for energies below the next four pion threshold but even beyond
that point it is approximately valid.

In the level of approximation considered in this work (large $N$ limit and
first order in
$m^2_{\pi}/F^2$), the dominant channel
is the $I=J=0$. As it can be expected the elastic unitarity condition is not
exact but
 it can be shown that one has
\begin{eqnarray}
Im a_{00}=
\sigma |a_{00}|^2+
O\left(\frac{1}{N}\right)+
O\left[\left(\frac{m^2_{\pi}}{F^2}\right)^2\right]
\end{eqnarray}
{}From the partial waves it is possible to define in the standard way the phase
shifts
$\delta$ as
\begin{equation}
t_{IJ}(s)= e^{i\delta_{IJ}(s) }\sin \delta_{IJ}(s)/\sigma(s)
\end{equation}
Now one can try to compare the results of our approximation with the
experiment. Then
one finds that  the  $I=J=0$ pion scattering phase shifts can be fitted with
the
 simple choice
 $G^R(s;\mu)=1$ and  $J^R(s;\mu)=1$ for $\mu=775 MeV$, as it is shown in
 Fig.3. Concerning  the $I=0,J=2$ channel (which is the other relevant for
the $\gamma\gamma \rightarrow \pi^0\pi^0$ reaction), the prediction of the
large $N$
 approximation is that
  it is suppressed at leading order in the $1/N$ expansion. In any case, one
can use
the above  equations with the $\mu$ value fitted for the $a_{00}$ channel. The
result,
 which is in fact
a prediction, is
also shown in Fig.3 and,  as it can be seen, the agreement with the
experimental data
is pretty good.

At the same level of approximation, the diagrams contributing
to the ${\gamma}{\gamma} \rightarrow {\pi}^{0}{\pi}^{0}$ reaction can be
 found in Fig.2.b since it is not difficult to see that any other diagram is
suppressed by
 extra $1/N$ or $m^2_{\pi}/F^2$ powers.  Thus
the amplitude corresponding to this reaction can be written as
\begin{equation}
{\cal M}_{{\mu}{\nu}} = e^2\left(2g_{\mu\nu}I(s)-I(k_1k_2)_{{\mu}{\nu}}
-I'(k_1k_2)_{{\mu}{\nu}}\right)NA(s)
+O(1/N)+O((\frac{m^2_{\pi}}{F^2})^2)
\end{equation}
where
\begin{eqnarray}
I(s) & = & \int d \tilde l\frac{1}{(l-m^2_{\pi})[(l-p_1-p_2)^2-m^2_{\pi}]}
\\ \nonumber
I(k_1k_2)_{{\mu}{\nu}}& = &\int d \tilde l\frac{(2l+k_1)_{\mu}(2l-k_2)_{\nu}}
{(l-m^2_{\pi})[(l+k_1)^2-m^2_{\pi}][(l-k_2)^2-m^2_{\pi}]}
\\ \nonumber
I'(k_1k_2)_{{\mu}{\nu}}& = &\int d \tilde l\frac{(2l-k_1)_{\mu}(2l+k_2)_{\nu}}
{(l-m^2_{\pi})[(l+k_2)^2-m^2_{\pi}][(l-k_1)^2-m^2_{\pi}]}
\end{eqnarray}
The last two integrals  have the same value, so that we
can write
\begin{equation}
{\cal M}_{{\mu}{\nu}} = 2e^2NA(s)
\int d\tilde{\ell} \frac{g_{{\mu}{\nu}}
 \left(l^{2}-m_{\pi}^{2}\right)- \left(2{\ell}+k_{1} \right)_{\mu}
 \left(2{\ell}-k_{2} \right)_{\nu}}{ \left({\ell}^{2}-m_{\pi}^{2} \right)
 \left[ \left({\ell}+k_{1} \right)^{2}-m_{\pi}^{2} \right]
\left[ \left({\ell}-k_{2} \right)^{2}-m_{\pi}^{2} \right]}
+O(1/N)+O((\frac{m^2_{\pi}}{F^2})^2)
\end{equation}
Note that this is the same integral that appears in the one-loop computation
in eq.14. Therefore we also find, in the approximation here considered,
the factorization property that we
 found there. In particular, eq.16 can be applied again although now the lowest
order ${\pi}^{+}{\pi}^{-}
\rightarrow {\pi}^{0}{\pi}^{0}$ cross-section has to be replaced  by
the cross-section obtained from eq.28, which is given by:
\begin{eqnarray}
{\sigma} \left({\pi}^{+}{\pi}^{-} \rightarrow {\pi}^{0}{\pi}^{0} \right) =
\frac{\mid A(s) \mid^2}{32{\pi} \left(NF^{2} \right)^{2}s}
\end{eqnarray}
The results obtained by using this formula, with the previously fitted value
for $\mu$
in the $I=J=0$ elastic pion scattering channel, are given in Fig.4. For
completeness we
also show the results obtained when the $J=0,I=2$ channel contribution is
included
by using the exact formula
\begin{eqnarray}
{\sigma} \left({\pi}^{+}{\pi}^{-} \rightarrow {\pi}^{0}{\pi}^{0} \right) =
\frac{8\pi}{9(s-4m^2_{\pi})}\sin^2(\delta_{00}-\delta_{20})
\end{eqnarray}
together with eq.16 and the phase shifts computed in the large $N$
approximation described above
which are shown in Fig.3. As it can be seen, the effect of the $I=2$ channel
is not very large,
since it is not so strongly interacting as the $I=0$. In any case our
computations
are compatible with the experimental data  in both cases (with and without the
introduction of
the $I=2$ channel) if one takes into account the large error bars.
Thus, the large $N$ approximation makes  it possible to fit all the data
shown in Fig.3 and
 Fig.4 with just one parameter, namely the scale $\mu = 775 MeV$.

\section{Discussion}

In some sense the aim
 of this work is to clarify to what extent the above mentioned factorization
relation
between the  ${\gamma}{\gamma} \rightarrow {\pi}^{0}{\pi}^{0}$ and the
$\pi^+\pi^- \rightarrow \pi^0\pi^0$ amplitudes occurs beyond the one-loop
 approximation. Note that, as it was stressed by Pennington and Morgan in [12],
 the
computation of the exact amplitude
will require the knowledge of the $\pi^+\pi^- \rightarrow \pi^0\pi^0$ amplitude
not only on-shell but also off-shell in order to make loops. In addition,
the partial waves of the first reaction behave
 at threshold as $s^{J/2}(s-4m^2_{\pi})^{J/2}$ whereas the second goes
as $(s-4m^2_{\pi})^{J}$ and therefore the factorization formula cannot be
correct.
 Moreover, from
elastic unitarity (at the lowest order in the electromagnetic interactions),
we have
the relation
\begin{equation}
 Im{\cal F}^{\Lambda}_{IJ}=\sigma {\cal F}^{\Lambda*}_{IJ} t_{IJ}
\end{equation}
involving the ${\gamma}{\gamma} \rightarrow {\pi}^{0}{\pi}^{0}$ and the
$\pi\pi$ partial wave (we have used the standard definitions which can be
found for
 instance in [5]). In this reference it was shown how it is possible to use
dispersion
 relations to improve the unitary behaviour of  one-loop $\chi PT$, both for
elastic
pion scattering and $\gamma\gamma \rightarrow \pi^0\pi^0$. As a result, an
excellent agreement
 with the data can be found in both cases. These
 dispersion relations have to be applied to each channel independently.
Therefore, as the
 total
amplitude is a linear combination, with well defined coefficients, of the
different
channel partial waves, we arrive again to the conclusion that the
factorization relation
cannot be hold in general.

Therefore, for many different reasons, we known that the factorization formula
which appears
as a result of the large $N$ approximation (as well as  the original one-loop)
cannot be true for the exact amplitudes. Then the natural question arises: Is
there anything wrong with
the large $N$ approximation? and in that case: Why does it reproduce so well
the experimental data?

Of course the answer to the first question is no. The large $N$ approximation
is a well
defined expansion in the $1/N$ parameter and it is perfectly right. In
particular
the computations done by using this approximation satisfy the current
constrains
of any Quantum Field Theory like unitary, analyticity, crossing and so on,
modulo higher
order corrections in the expansion parameter, i.e. $1/N$ (see eq.33), exactly
as it happens in any
other perturbative expansion. The reason why it gives rise to the
factorization formula
is the following: according to eq.27 the $A$ function, which in principle
depends on
$s,t$ and $u$, is only a function of $s$ at leading order in the $1/N$
expansion.
In addition, from eq.30 we see that, strictly speaking, only the $I=0$ channel
contributes
in this approximation. In other words we can say that to leading order in the
large $N$ expansion
the only non-zero channel is  $I=J=0$. Then the
$\pi^+\pi^- \rightarrow \pi^0\pi^0$ amplitude only
depends on $s$ and only has an isoscalar component. This explains why  all the
previous mentioned
objections to the factorization formula do not apply to this case as it
happens in the
general case where more than one channel contributes to the reaction. Therefore
 there is no
inconsistency in the factorization formula obtained in the large $N$ limit.
However,
this formula should be corrected by higher order computations (both in
$1/N$ and in $m^2_{\pi}/F^2$) which will also include other channels different
from the $I=J=0$.
In that case the factorization formula is not justified any more.

Of course one can ask  how good  the large $N$ approximation
to leading order in $1/N$ and in $m^2_{\pi}/F^2$ can be, since it only
 includes the $I=J=0$ channel contribution. The  answer
can be found in Fig.4. There we see that the agreement with the experimental
data
is good, and in fact it considerably improves the one-loop computation (the
two-loop computation
 in [5]
also reproduces well the data but depends on more parameters)

\section{Conclusions}

The main results of our work are the following:

In order to properly define the large $N$ expansion for two flavor $\chi PT$ we
have parametrized the coset space $S^3 \rightarrow S^N$ by using standard
coordinates
instead of  exponential (chiral) coordinates. We have reobtained the well
known one-loop results for
the $\gamma\gamma \rightarrow \pi\pi$ reactions with this parametrization in
a much simpler
way, thus illustrating, with a very non-trivial example, the $S$ matrix
invariance under
reparametrizations of the NLSM.

We have computed the $\gamma\gamma \rightarrow \pi^0\pi^0$ scattering
amplitude to leading
order in the large $N$ approximation up to  order $m^2_{\pi}/F^2$. In this
approximation
 this amplitude turns out to  be
proportional to that of $\pi^+\pi^- \rightarrow \pi^0\pi^0$   (the
factorization
 formula) according to  Im's conjecture (but in principle only for small
$m^2_{\pi}/F^2$).

Using a one-parameter fit of the elastic pion scattering in the $I=J=0$
 channel obtained
in the same approximation we reproduce, as a prediction, both the
 $I=2,J=0$ pion phase shift and the
$\gamma\gamma \rightarrow \pi^0\pi^0$
cross-section experimental data (see Fig. 1 and 2).

{}From very general considerations it can be shown  that the factorization
formula
 cannot be exact
and this was the basis for the  Morgan and Pennington criticism to  Im's
results. However
Im's result is correct in the framework of the large $N$ approximation (for
small $m^2_{\pi}/F^2$)
since only the $I=J=0$ channel contributes to the leading order in the $1/N$
expansion.
As a matter of fact, as the $\gamma\gamma \rightarrow \pi^0\pi^0$ is dominated
by this
channel, the factorization formula yields a good approximation as it was
suggested
by Donoghue et al in [3]. This would not be the case if, for example, the
$\rho$ resonance
played an important role in the reaction, since it cannot be reproduced by the
leading order
large $N$ approximation [11].

In summary, we consider that the large $N$ approach to $\chi PT$ is a very
interesting
complementary alternative to the standard one-loop computations which can be
useful in
many cases in order to describe the relevant physics of the considered
phenomena. This is
 particularly true
in those cases where the $I=J=0$ is dominant as it happens in the Higgs
physics.

\newpage

\vskip 1.0cm
{\bf Acknowledgments}  \\

This work has been supported in part by the Ministerio de Educaci\'on y Ciencia
(Spain) (CICYT AEN93-0776) and COLCIENCIAS (Colombia). The authors thank
F. Cornet,
 J.F. Donoghue,
J. Gasser and J. R. Pel\'aez for useful information and discussions.

\newpage

{\bf\large Figure Captions}

Fig. 1: Diagrams contributing to a) the $\gamma\gamma \rightarrow \pi^0\pi^0$
reaction at  one-loop level, b)  the tree level $\gamma\gamma
 \rightarrow \pi^+\pi^-$, c)
  the one-loop $\gamma\gamma \rightarrow \pi^+\pi^-$ and
d)   the one-loop pion wave-function renormalization.

Fig.2: a) Diagrams contributing to the $\pi-\pi$ elastic scattering at leading
order
in the $1/N$ expansion up to order $m^2_{\pi}/F^2$. The large black dot
represents the
 pion coupling proportional to $m^2_{\pi}$. b) Diagrams contributing to the
 $\gamma\gamma \rightarrow
 \pi^0\pi^0$ reaction to the same level of approximation. The black square
represents the
addition of all the diagrams appearing in a).

Fig 3: Phase shifts for $J=0$ elastic pion scattering. The dashed line
 represents the fit done with the leading order of the $1/N$ expansion up to
the order
 $m^2_{\pi}/F^2$ for the $I=0$ channel and the prediction for the $I=2$
channel.
The continuous lines are the phase shifts as obtained from the standard
one-loop $\chi PT$
 (see [11] for the details). The experimental
 data corresponds to: $\bigtriangleup$ ref.[15],
$\bigcirc$ ref.[16],$\Box$ ref.[17],$\diamondsuit$ ref.[18],
$\bigtriangledown$ ref.[19],$\star$ ref.[20],$\times$ ref.[21],$\bullet$
ref.[22].

Fig 4: Cross section for the process $\gamma \gamma \rightarrow \pi ^{0}
\pi ^{0} $ with $\mid  cos\theta \mid \leq 0.8$. The continuous line represents
the prediction of leading order of the $1/N$ expansion up to the order
 $m^2_{\pi}/F^2$. The dashed line represents the same but including the
 contribution of the
$I=2$ channel. The experimental data comes from Cristal Ball [1].

\end{document}